\documentclass[letterpaper]{jpconf}
\usepackage{graphicx}
\usepackage{multicol}
\usepackage{multirow}
\begin{document}
\title{$\pi^{0}$ and the T2K Experiment}

\author{Benjamin Still\footnote{Supported by the EU grant No.~207282-T2KQMUL}, on behalf of the T2K collaboration.}

\address{Queen Mary, University of London\\b.still@qmul.ac.uk}

\begin{abstract}
The Tokai to Kamioka neutrino oscillation experiment aims to
determine the third and final lepton mixing angle $\theta_{13}$, through
a measurement of the sub-dominant oscillation
$\nu_{\mu}\rightarrow\nu_{e}$. The oscillation is a maximum
as $\nu_{\mu}$ travel 295~km from the near detectors to the far
Super-Kamiokande detector. Single $\pi^{0}$ production from neutral 
current (NC1$\pi^{0}$) neutrino interactions are a significant
background to $\nu_{e}$ events in water Cherenkov detectors such as
Super-Kamiokande. To reduce this background, the off-axis ND280 near
detector contains 
water target regions to determine the cross-section of such
NC1$\pi^{0}$ events. Here, I discuss the status of the detectors and
tools in preparation for physics data taking. 
\end{abstract}

\section{Introduction}
Neutrino oscillation, first confirmed by the Super-Kamiokande (Super-K)
experiment in 1998\cite{Fukuda:1998mi}, is a young field of particle 
physics. It is about to enter the age of precision measurements with a
new generation of neutrino beam and reactor experiments closing in on
the third and final unmeasured lepton mixing angle, $\theta_{13}$. 

\begin{figure}[!bh]
 \begin{center}
  \begin{tabular}{cc}
   \begin{minipage}{0.5\textwidth}       
    \includegraphics[keepaspectratio=true,width=\textwidth]{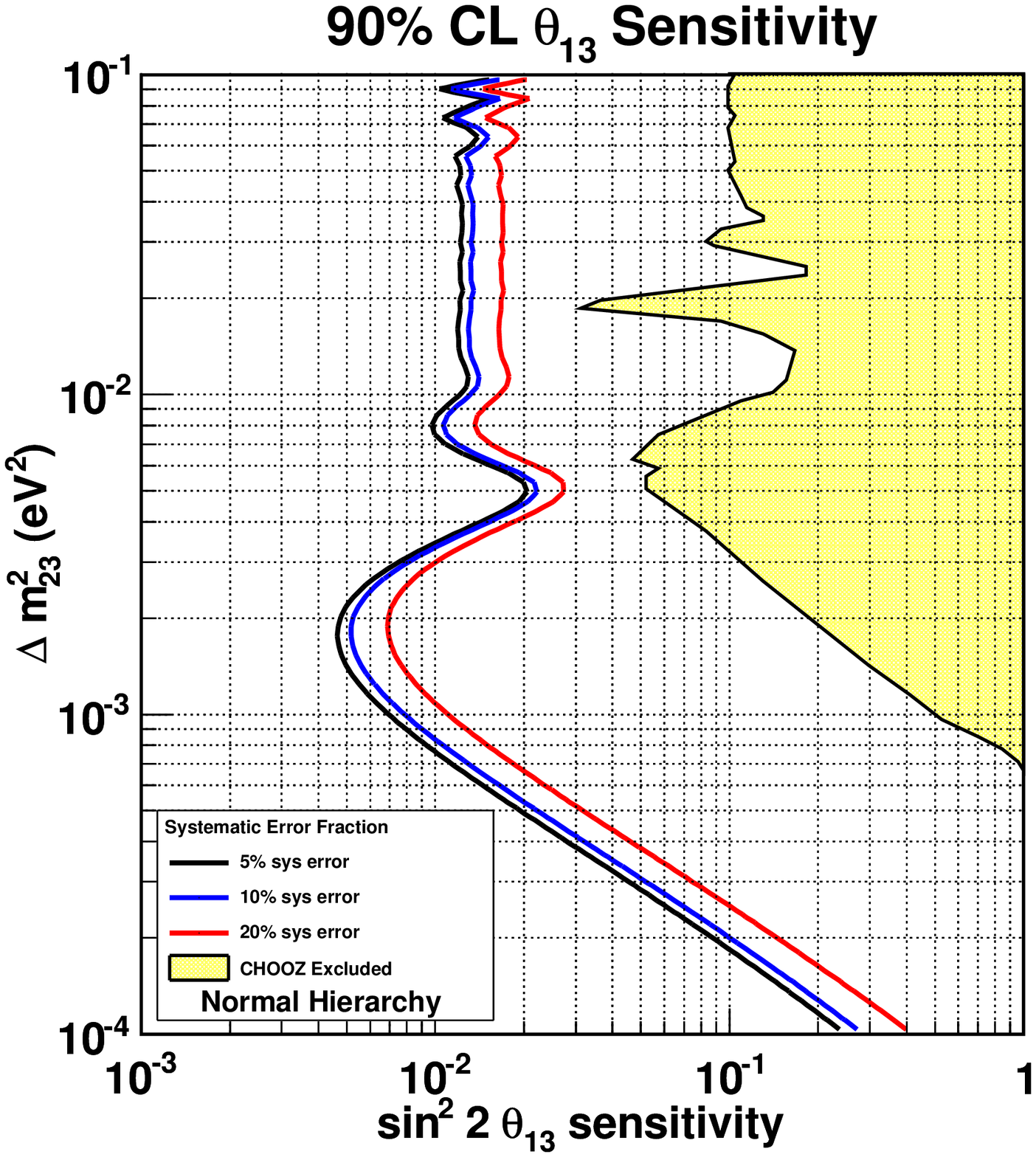}
   \end{minipage}
   &    
   \begin{minipage}{0.5\textwidth}
     \begin{center}
       \begin{tabular}{|l|l|l|}
         \hline
         Systematic Source & Limit & Detectors\\
         \hline
         \hline
         Beam & $<1$~mrad& INGrid\\
         Direction & & ND280\\
          & & MuMon\\
         \hline
         $\nu$ Energy & $<10$~\% & ND280\\
         Spectrum & & NA61\\
          & & INGrid\\
         \hline
         Beam $\nu_{e}$ & $<10$~\% & ND280\\
         Component & (relative) & NA61\\
         \hline
         NC1$\pi^{0}$ & $<10$~\% & ND280\\
         x-sec & & \\
         \hline
       \end{tabular}
     \end{center}
   \end{minipage}
   \\
   (a) & (b)
  \end{tabular}
    \caption{(a) T2K sensitivity to $\theta_{13}$ at the 90\% confidence level 
      as a function of $\Delta m^{2}_{23}$.  Beam is assumed to 
      be running at 0.75~MW for $5\times\,10^{7}$~s, using the 22.5~kton 
      fiducial volume SK detector.  5\%, 10\% and 20\% 
      systematic error fractions are plotted.  The yellow region 
      has already been excluded to 90\% confidence level by the 
      Chooz reactor experiment.  The following oscillation 
      parameters are assumed: $\sin^{2}2\theta_{12} = 0.8704$, 
      $\sin^{2}2\theta_{23} = 1.0$, 
      $\Delta m^{2}_{12} =  7.6\times10^{-5}$~{eV}$^{2}$, 
      $\delta_{CP}=0$, normal hierarchy.
      (b) T2K target systematic errors and near detectors involved.}\label{sens_syst:fig}
 \end{center}
\end{figure}

The
current direct upper limit of this small mixing angle is set by the CHOOZ
reactor experiment, 90\%~C.L. shown as the shaded region of
Fig.~\ref{sens_syst:fig}(a)  
$\sin^{2}2\theta_{13}<0.19$\cite{Apollonio:1999ae} (at $\Delta
m^{2}_{23}=1.9\times 10^{-3}$~eV), which is consistent
with zero within errors. There is a hint, however, of a non-zero value for
$\theta_{13}$\cite{Fogli:2008jx} from global fits to all neutrino
oscillation data; solar, atmospheric, reactor and accelerator. If
$\theta_{13}\neq 0$ then neutrino oscillation 
experiments will have the task of determining the level of CP
violation in the lepton sector.

The Tokai to Kamioka (T2K) experiment is the first of the new
generation of super-beam neutrino oscillation experiments. T2K uses
the worlds most
powerful man-made beam of neutrinos, generated at the Japan Proton
Accelerator Research Center (J-PARC) in Tokai, Ibaraki Prefecture,
Japan. T2K will have a sensitivity to $\sin^{2}2\theta_{13}$ an order
of magnitude better than the CHOOZ limit. The beam
is sampled by a suite of near detectors at 280~m from 
the beam target station; profiling the beam composition, direction and energy
spectrum. Different detectors within the suite aim to constrain errors
on key systematic sources to within 10\%, Fig.~\ref{sens_syst:fig}(b).
The neutrinos then travel 295~km across Japan to be sampled
again by the refurbished and upgraded Super-Kamiokande
detector.

\section{Measuring $\theta_{13}$: $\nu_{\mu}\rightarrow\nu_{e}$}
The T2K neutrino beam is composed of more than $99\%$ $\nu_{\mu}$, the
majority of which oscillate to $\nu_{\tau}$ over the 295~km journey
thanks to a near maximal mixing 
$\theta_{23}$. The sub-dominant oscillation of $\nu_{\mu}$ to $\nu_{e}$
is driven by the $\theta_{13}$ mixing angle, as seen in the
simplified three neutrino oscillation probability relevant to T2K:
\begin{eqnarray}
P\left(\nu_{\mu}\rightarrow\nu_{e}\right)&\simeq
&\sin^{2}2\theta_{13}\sin^{2}\theta_{23}\sin^{2}\Delta\nonumber\\
&&\pm\alpha\sin 2\theta_{13}\cos\theta_{13}\sin\delta\sin2\theta_{12}\sin2\theta_{23}\sin^{3}\Delta\nonumber\\
&&-\alpha\sin 2\theta_{13}\cos\theta_{13}\cos\delta\sin2\theta_{12}\sin2\theta_{23}\cos\Delta\sin 2\Delta\nonumber\\
&&+\alpha^{2}\cos^{2}2\theta_{23}\sin^{2}2\theta_{12}\sin^{2}\Delta,
\end{eqnarray}
where $\alpha\equiv\Delta m_{21}^{2}/\Delta m_{31}^{2}$ and
$\Delta=\Delta m_{31}^{2}L/4E$.

The neutrino energy spectra are determined at both near and far detector
from
Charged Current Quasi-Elastic (CCQE) events. These events are selected
with high efficiency and are theoretically well understood; a
tree level two body interaction, with just a recoil nucleon and
relevant charged lepton in the final state. 

There are two major backgrounds to $\nu_{e}$
appearance at Super-K. The first is the small intrinsic $\nu_{e}$
content of the beam produced at J-PARC, discussed elsewhere in these
proceedings~\cite{Melissa:LLWI}. The second arise from the 
production of a lone $\pi^{0}$ from neutral current $\nu_{\mu}$
interaction (NC1$\pi^{0}$), discussed in the next section.

\section{NC1$\pi^{0}$ Background}
Super-K has excellent discrimination between minimally ionizing (MIP) and
electro-magnetically (EM) showering particles. This allows accurate
determination of charged current $\nu_{\mu}$ and $\nu_{e}$ event rates
respectively. CCQE signals require the detection of a single particle
of either MIP or EM showering type.

A $\pi^{0}$ decays into two photons, which should show
as two EM showering type particles in Super-K. Issues arise if just a
single photon is observed, producing a signature indistinguishable
from a single
electron, a signature of a $\nu_{e}$ CCQE interaction. This may occur if one photon is
of insufficient energy to pair produce particles above the Cherenkov
threshold of Super-K. Such background may also arise if both photons
signals overlap posing as just a single EM showering particle. 

\begin{figure}[!h]
 \begin{center}
  \begin{tabular}{cc}
   \begin{minipage}{0.4\textwidth}       
    \includegraphics[keepaspectratio=true,width=1.2\textwidth]{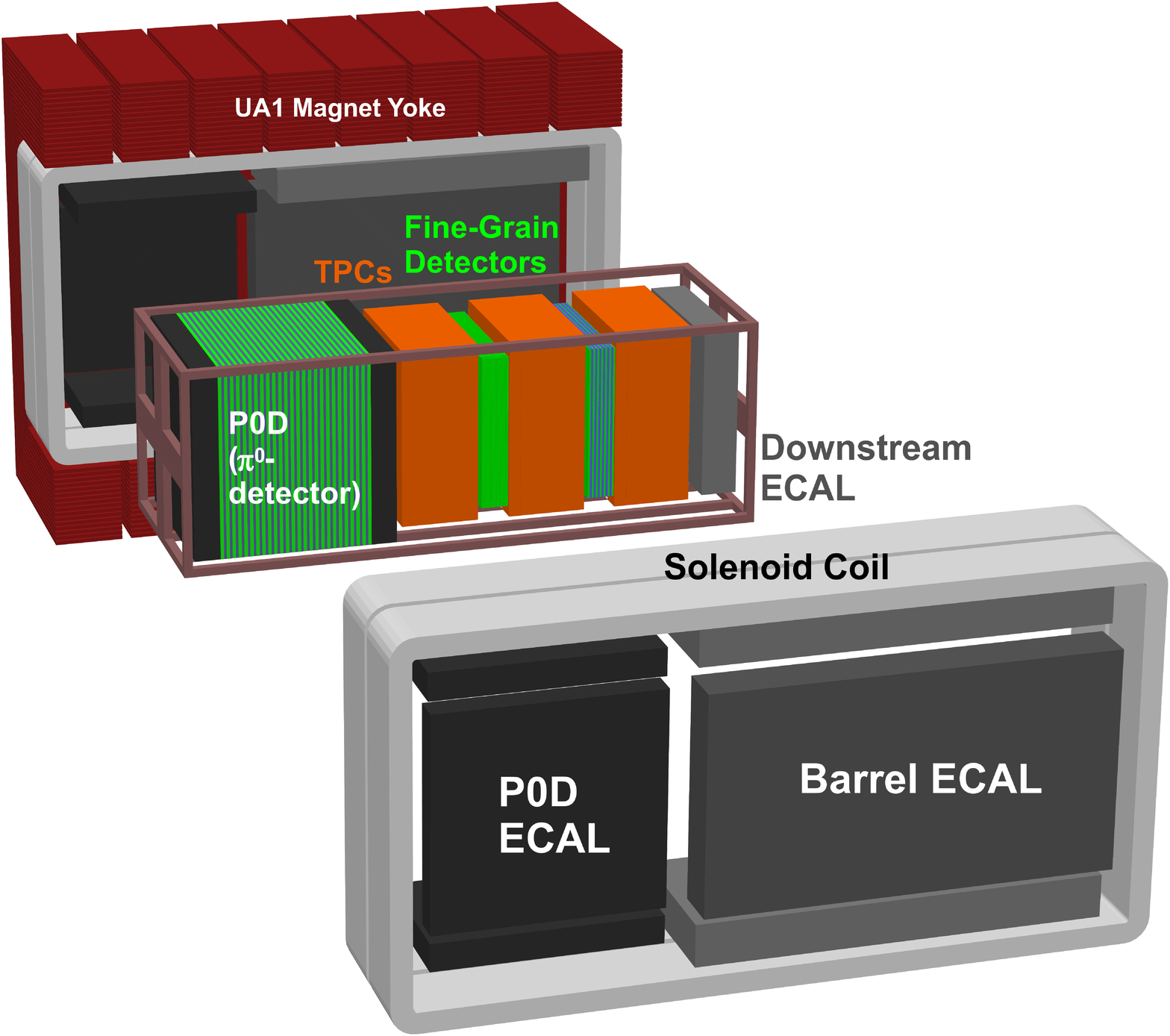}
   \end{minipage}
   &    
   \begin{minipage}{0.6\textwidth}
    \includegraphics[keepaspectratio=true,width=1.2\textwidth]{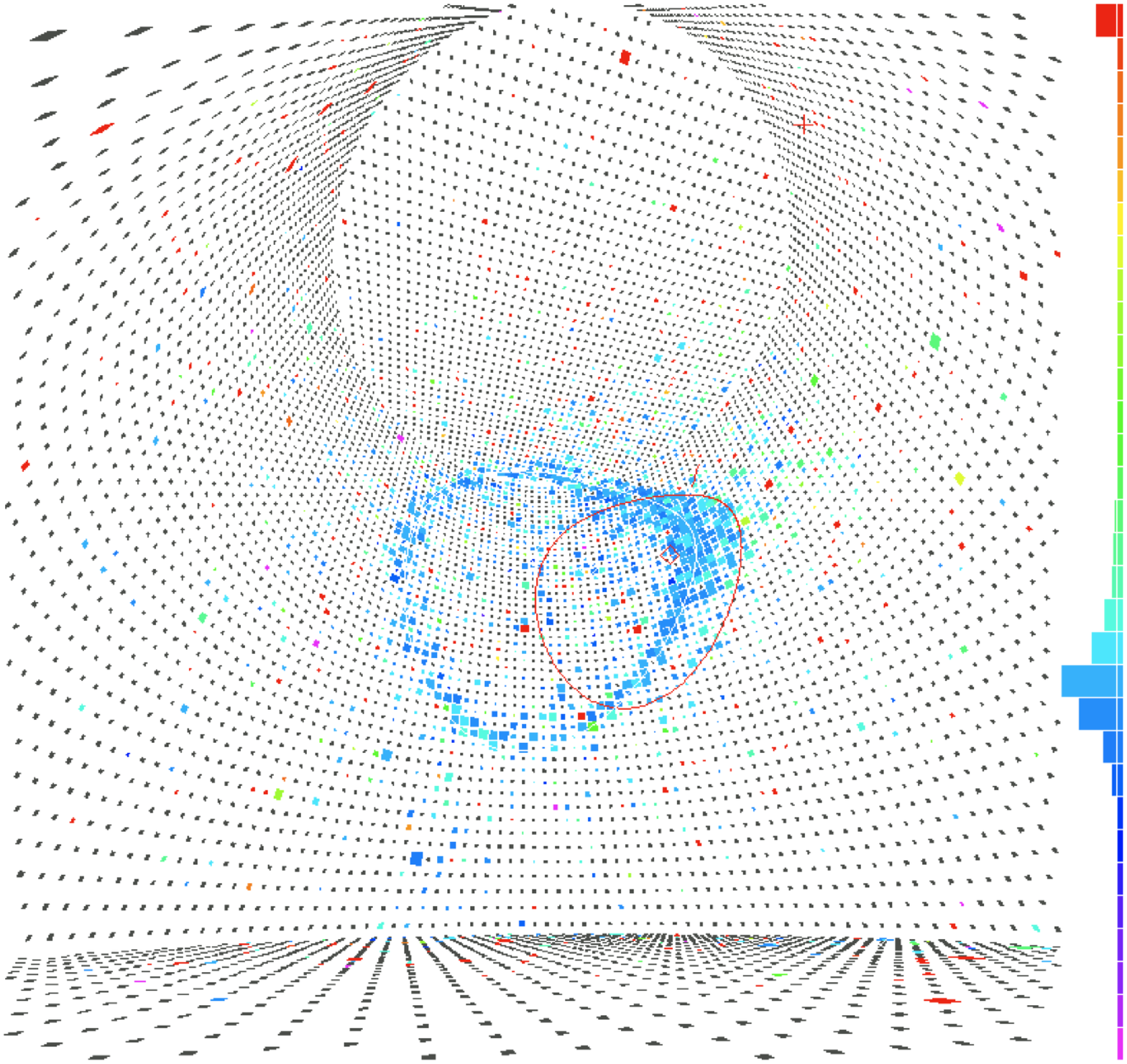}
   \end{minipage}
   \\
   (a) & (b)
  \end{tabular}
    \caption{(a) Schematic of the ND280 near detector. (b) A recovered
    single $\pi^{0}$ from the K2K neutrino beam~\cite{tscan:web}, with
    second ring fitted.}\label{nd280_tscan:fig}
 \end{center}
\end{figure}

Super-K has a collection of algorithms in place to minimise the
latter: a second Cherenkov ring search and invariant mass
reconstruction, as seen in Fig.~\ref{nd280_tscan:fig}(b). Despite this
reduction, the remaining events still pose a significant background to
the small expected number of oscillated $\nu_{e}$ events.

Over the projected lifetime of the T2K experiment (0.75~MW\,$\times5\,\times10^{7}$~s), assuming $\sin^{2}\theta_{13}=0.1$, just below 
that of the current limit, we expect 143 $\nu_{e}$ appearance signal
events at Super-K, see Table~\ref{nue_exp:table}. As we approach the
sensitivity limit of the experiment, the number of signal events
scales almost linearly, so just 14 at a value of
$\sin^{2}\theta_{13}=0.01$.  

\begin{table}[!t]
\begin{center}
\begin{tabular}{|c|c|c|c|c|}
\hline
\hline
\multirow{2}{*}{$\sin^{2}\theta_{13}$} & \multicolumn{3}{c|}{Backgrounds} & \multirow{2}{*}{Signal}\\
 & $\nu_{\mu}$ Induced & Beam $\nu_{e}$ & Total BG & \\
\hline
\hline
0.1 & 10 & 16 & 26 & 143\\
\hline
\end{tabular}
\caption{Expected reconstructed events at Super-K for the projected
  lifetime of the T2K experiment, 0.75~MW\,$\times5\,\times10^{7}$~s.} \label{nue_exp:table}
\end{center}
\end{table} 

Over the same lifetime a total of 26 irreducible
background events are expected; 10 from $\nu_{\mu}$ induced NC1$\pi^{0}$
events and 16 expected intrinsic beam $\nu_{e}$, Table~\ref{nue_exp:table}. One can see,
therefore, that the sensitivity of T2K to the value of $\theta_{13}$
is heavily dependent on our understanding of the $\nu_{e}$
backgrounds. It is the goal of the ND280 off-axis near detector to measure the
cross-section of NC1$\pi^{0}$ and intrinsic beam $\nu_{e}$.

\section{Measuring $\pi^{0}$ at ND280}
The ND280 off-axis near detector is a collection of specialised sub-detectors
all housed within the refurbished UA1 magnet which provides a uniform
dipole field of 0.2~T. The downstream tracker region, consisting of
time projection chambers (TPCs) and finely grained plastic scintillator
detectors (FGDs) are discussed further elsewhere~\cite{Kendall:LLWI}. 

In the most upstream
end of the ND280 inner basket of the detectors is the Pi-Zero
Detector (P0D), a plastic scintillator tracking detector. The P0D has a
large fiducial mass, which also include bladders of water, to yield large
numbers of neutrino interactions. Triangular scintillator bars provide
accurate shower pointing. The bladders of inactive water will be filled
or emptied at various stages of data taking, allowing the
determination of the relative event rate of NC1$\pi^{0}$ production on
water.

Over the same lifetime of the T2K experiment 
(0.75~MW\,$\times5\,\times10^{7}$~s) the P0D expects to successfully reconstruct and select over 9000
NC1$\pi^{0}$ events. With projected systematic uncertainties summing
to less than 8\% the P0D is envisaged to perform its task of
understanding the NC1$\pi^{0}$ interaction rate to better than the
10\% target uncertainty.

A complementary measurement of NC1$\pi^{0}$ will also be made by the
tracker region of the ND280, at reduced statistics. The
electromagnetic calorimeters (ECals) surround the tracker region,
catching and containing escaping photons from decaying 
$\pi^{0}$. This complementary measurement will reduce the
associated systematic errors. 

\section{Status and Conclusions}

The hardware upgrade of Super-Kamiokande IV is complete and new techniques of
$\pi^{0}$ and electron discrimination are in development. 
All central sub-detectors of the ND280 near detector, including the P0D,
have been commissionned and are taking data in unison. The
electromagnetic calorimeters (ECals) surrounding the inner detector
regions will be installed this summer during a scheduled beam
shutdown, which will complete the ND280 detector.

The INGrid near detector is performing well, monitoring the beam
position to better than the required 1~mrad. The J-PARC neutrino
beamline itself is currently supplying protons to the target steadily
at around 40~kW.

Physics analyses are in heavy development in every area, including
$\pi^{0}$ event selection. First results of event rate and and vertex
positions are expected this summer 2010.

\section*{Bibliography}
\bibliographystyle{nar}
\bibliography{LLWIproceedings}






\end{document}